# Improving Efficiency of Training a Virtual Treatment Planner Network via Knowledge-guided Deep Reinforcement Learning for Intelligent Automatic Treatment Planning of Radiotherapy

Chenyang Shen[1,2*], Liyuan Chen[1], Yesenia Gonzalez[1,2], Xun Jia[1,2*]

1. Medical Artificial Intelligence and Automation (MAIA) Laboratory, Department of Radiation Oncology, University of Texas Southwestern Medical Center, Dallas, TX 75390, USA
2. innovative Technology Of Radiotherapy Computation and Hardware (iTORCH) Laboratory, Department of Radiation Oncology, University of Texas Southwestern Medical Center, Dallas, TX 75390, USA

*Corresponding Authors' emails: Chenyang.Shen@UTSouthwestern.edu and Xun.Jia@UTSouthwestern.edu

## Abstract

We previously proposed an intelligent automatic treatment planning framework for radiotherapy, in which a virtual treatment planner network (VTPN) is built using deep reinforcement learning (DRL) to operate a treatment planning system (TPS) by adjusting treatment planning parameters in it to generate high-quality plans. We demonstrated the potential feasibility of this idea in prostate cancer intensity-modulated radiation therapy (IMRT). Despite the success, the process to train a VTPN via the standard DRL approach with an $\epsilon$-greedy algorithm was time consuming. The required training time was expected to grow with the complexity of the treatment planning problem, preventing the development of VTPN for more complicated but clinically relevant scenarios. This study proposed a knowledge-guided DRL (KgDRL) approach that incorporated knowledge from human planners to guide the training process to improve the efficiency of training a VTPN. Using prostate cancer IMRT as a testbed, we first summarized a number of rules in the actions of adjusting treatment planning parameters of our in-house TPS. During the training process of VTPN, in addition to randomly navigating the large state-action space, as in the standard DRL approach using the $\epsilon$-greedy algorithm, we also sampled actions defined by the rules. The priority of sampling actions from rules decreased over the training process to encourage VTPN to explore new policy on parameter adjustment that were not covered by the rules. To test this idea, we trained a VTPN using KgDRL and compared its performance with another VTPN trained using the standard DRL approach. Both networks were trained using 10 training patient cases and 5 additional cases for validation, while another 59 cases were employed for the evaluation purpose. It was found that both VTPNs trained via KgDRL and standard DRL spontaneously learned how to operate the in-house TPS to generate high-quality plans, achieving plan quality scores of 8.82 (±0.29) and 8.43 (±0.48), respectively. Both VTPNs outperformed treatment planning purely based on the rules, which had a plan score of 7.81 (±1.59). VTPN trained with eight episodes using KgDRL was able to perform similarly to that trained using DRL with 100 epochs. The training time was reduced from more than a week to ~13 hours. The proposed KgDRL framework was effective in accelerating the training process of a VTPN by incorporating human knowledge, which will facilitate the development of VTPN for more complicated treatment planning scenarios.

## 1. Introduction

Inverse treatment planning of modern radiation therapy modalities, such as Intensity Modulated Radiation Therapy (IMRT) or Volumetric Modulated Arc Therapy (VMAT), is often achieved by solving an optimization problem. Objective functions of these optimization problems typically have multiple terms and constraints designed for various considerations, as well as a set of treatment planning parameters (TPPs) such as weighting factors, dose limits, and volume constraints. The values of these TPPs critically affect the resulting plan quality. While a treatment planning system (TPS) can solve the optimization problem for a given set of TPP values, human planners are still needed in the treatment planning process to determine the values of TPPs to achieve plans with clinically acceptable quality. The whole process with extensive interactions between a human planner and the TPS is time consuming and labor intensive. The resulting plan quality is affected by a number of human factors, such as the experience of the planner and the available planning time (Das *et al.*, 2008; Nelms *et al.*, 2012).

To solve this problem and fully automate the treatment planning process, a number of methods have been successfully developed, including greedy approaches (Xing *et al.*, 1999; Lu *et al.*, 2007; Wu and Zhu, 2001; Wang *et al.*, 2017), heuristic approaches (Yang and Xing, 2004; Wahl *et al.*, 2016; Yan and Yin, 2008), fuzzy inference (Yan *et al.*, 2003b; Yan *et al.*, 2003a; Holdsworth *et al.*, 2012; Holdsworth *et al.*, 2010), and statistics-based methods (Lee *et al.*, 2013; Boutilier *et al.*, 2015; Chan *et al.*, 2014). More recently, deep learning based methods (Shen *et al.*, 2020b) have shown their great promise in the context of automatic treatment planning (Nguyen *et al.*, 2020; Nguyen *et al.*, 2019; Shen *et al.*, 2019a; Fan *et al.*, 2019; Mahmood *et al.*, 2018; Shen *et al.*, 2020a). In particular, deep reinforcement learning (DRL) has been employed to develop an intelligent automatic treatment planning framework. Within this framework, a virtual treatment planner network (VTPN) was built to model the intelligent behaviors of human planners in the treatment planning process. Trained via an end-to-end DRL process, the VTPN was able to operate a TPS by adjusting the TPPs in it to generate high-quality plans. Specifically, similar to the human planner's role in treatment planning, the VTPN repeatedly took a state of the optimization problem as input, e.g. the dose-to-volume histogram (DVH) of a plan generated by the optimization engine under a given set of TPPs, and determined an action to adjust the TPPs to improve the resulting plan quality. The feasibility of this approach has been demonstrated in preliminary studies in exemplary problems of high-dose-rate (HDR) brachytherapy for cervical cancer (Shen *et al.*, 2019a) and IMRT for prostate cancer (Shen *et al.*, 2020a).

Despite the initial success, a major concern was low efficiency of training a VTPN. Training a VTPN requires a large number of training data in the form of state-action pairs, i.e. the combinations of plan DVHs and corresponding actions of adjusting TPPs. The standard DRL approaches employ an $\epsilon$-greedy algorithm to navigate the state-action space and generate the training data (Mnih *et al.*, 2015; Silver *et al.*, 2016; Silver *et al.*, 2017; Shen *et al.*, 2019a; Shen *et al.*, 2019b; Shen *et al.*, 2018; Shen *et al.*, 2020a). Specifically, at each training step, it selects the optimal action predicted by the current VTPN for the plan state with a probably of $(1 - \epsilon)$, and a random action among all possible actions with a probably of $\epsilon$. The parameter $\epsilon$ is usually set to be close to unity in the early stage of training and is gradually reduced over the training process. The purpose of this strategy is to allow a random exploration of the state-action space initially, and to progressively focus on those actions made by the trained VTPN. Generally speaking, it is necessary for the VTPN to observe both appropriate actions, i.e. those able to improve the plan quality, as well as inappropriate ones, so that the VTPN can conclude proper behaviors and be trained to learn the desired ones. However, as can be imagined, it is quite challenging for the $\epsilon$-greedy algorithm to collect a sufficient amount of data with appropriate actions by the random exploration of the state-action space due to the scarceness of those appropriate actions as compared to those inappropriate ones. As a consequence, it often requires a large number of training steps for the $\epsilon$-greedy algorithm to collect sufficient samples to successfully train the VTPN.

The second hurdle substantially reducing computational efficiency was related to solving the optimization problem. Different from other DRL applications, such as playing Atari games or the game of Go (Mnih *et al.*, 2015; Silver *et al.*, 2016; Silver *et al.*, 2017), where the response to an action, e.g. score of a move in the Atari games, can be obtain in almost real-time, it takes much longer time to evaluate the influence in plan quality caused by an adjustment on the TPPs. The change in plan quality can only be computed by comparing two plans prior to and after the TPP adjustment, for which plan optimization using the adjusted TPPs has to be performed. Given the fact that DRL usually requires a huge number of steps to navigate and sample the state-action space, the training time of VTPN can be days or even weeks in previous proof-of-principle studies using in-house TPSs (Shen *et al.*, 2019a; Shen *et al.*, 2020a), in which only a small number of possible actions existed in those TPSs. It is expected that the efficiency would become a more severe concern, when extending the DRL-based VTPN methods to more sophisticated but clinically relevant scenarios, e.g. having the VTPN to adjust TPPs of a clinically realistic TPS. In these scenarios, the much larger state-action space due to the significantly more number of adjustable TPPs, as well as the longer time required to solve the treatment planning optimization problems would considerably prolong the training process, potentially rendering the VTPN method impractical.

In this paper, we propose a knowledge-guided DRL (KgDRL) scheme that integrates general experience in TPP adjustment from human planners with the standard $\epsilon$-greey algorithm to guide the navigation process in the state-action space, and hence improve the training efficiency of VTPN. Given rules of TPPs adjustment summarized based on human experience, training of VTPN will have a large chance to sample proper actions for states, making KgDRL more efficient and effective than that of the original DRL which purely relies on the $\epsilon$-greedy search. Meanwhile, the $\epsilon$-greedy search mechanism is still preserved in KgDRL, allowing VPTN to explore new policy of TPP adjustment that is not covered by the input human knowledge. Similar to (Shen *et al.*, 2020a), we will use the prostate cancer IMRT treatment planning problem as a testbed to study the KgDRL framework. We will analyze the performance of KgDRL and make comprehensive comparisons with the standard DRL to demonstrate its effectiveness in training the VTPN for intelligent automatic treatment planning.

## 2. Methods and Materials

### 2.1 Optimization engine and adjustable treatment planning parameters

Similar to (Shen *et al.*, 2020a) , our goal in this study was to train a VTPN to operate an in-house developed TPS by adjusting TPPs in the optimization engine to produce high-quality plans. The inverse plan optimization engine in the TPS solved the following fluence map optimization problem:

$$\min_x \frac{1}{2} \left\| Mx - d_p \right\|_-^2 + \frac{\lambda}{2} \left\| Mx - d_p \right\|_+^2$$

$$+ \frac{\lambda_{bla}}{2} \left\| M_{bla}x - \tau_{bla}d_p \right\|_+^2 + \frac{\lambda_{rec}}{2} \left\| M_{rec}x - \tau_{rec}d_p \right\|_+^2 , \qquad (1)$$

$$\text{s.t.} \, x \geq 0, \, D_{95\%}(Mx) = d_p.$$

$\| \cdot \|_-$ and $\| \cdot \|_+$ are $l_2$ norms computed for only negative and positive elements, respectively. $x \geq 0$ gives the beam fluence map to be determined, while $M$, $M_{bla}$, and $M_{rec}$ indicate the dose deposition matrices for planning target volume (PTV), bladder, and rectum, respectively. $d_p$ denotes prescription dose. The hard constraint $D_{95\%}(Mx) = d_p$ required that 95% of the PTV received dose no lower than the prescription dose. $d_p = 79.2$ Gy in this study.

There were five adjustable TPPs in this model, including the weighting factors $\lambda$, $\lambda_{bla}$, and $\lambda_{rec}$ to penalize overdose to PTV, bladder, and rectum, and the dose limits $\tau_{bla}$ and $\tau_{rec}$ to adjust dose to bladder and rectum. With a given set of TPPs, this optimization problem was solved using alternating direction method of multipliers (ADMM) (Boyd *et al.*, 2011; Glowinski and Le Tallec, 1989).

### 2.2 Virtual treatment planner network

We used a VTPN to automate the treatment planning process. Similar to the behavior of a human planner in treatment planning, given a plan, the VTPN decided a TPP adjustment action to modify the TPPs. The optimization engine was then launched using the updated TPPs to generate a new plan. This process continued, until a satisfactory plan was achieved or the maximal number of TPP adjustment steps was reached (Fig. 1(a)).

Specifically, in our formulation, the VTPN observed the DVH as the representation of a plan generated by solving the optimization problem in Eq. (1) under a given set of TPPs. The output of the VTPN were values of predicted quality of the plan for each action. Once the VTPN was determined, it can be used to decide an action for an input plan by selecting the action with the highest output value. VTPN was essentially an approximation of the optimal action-value function in the Q-learning framework (Watkins and Dayan, 1992). In many real applications including ours, the general form of such an optimal action-value function is unknown, and hence it is commonly parametrized via a DNN architecture possessing high flexibility and capacity to approximate complicate functions (Mnih *et al.*, 2013; Mnih *et al.*, 2015).

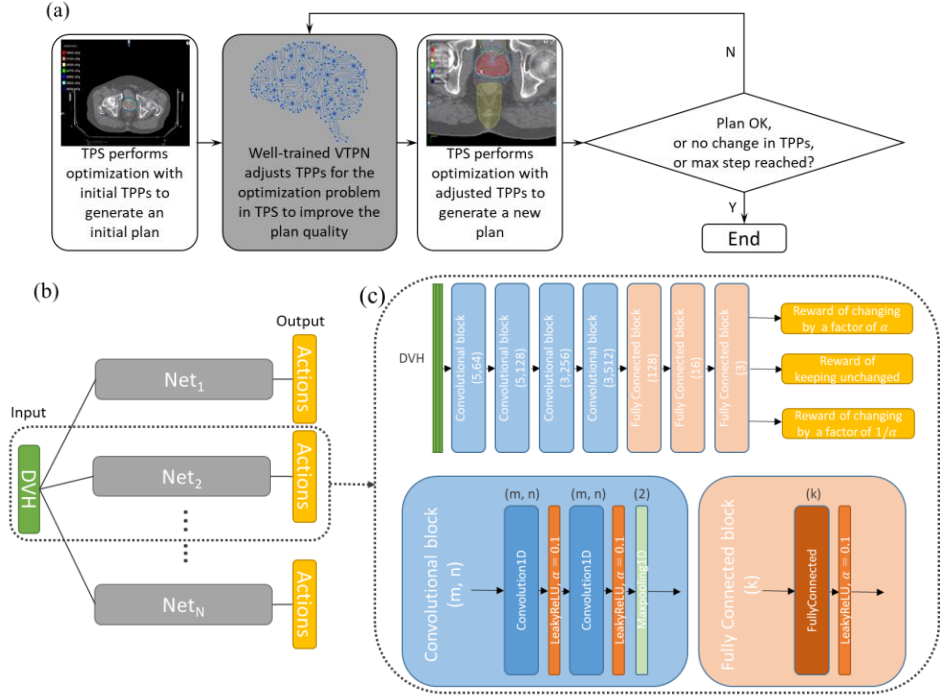

**Figure 1.** (a) The workflow of intelligent automatic treatment planning via VTPN. (b) The overall structure of the VTPN. (c) Detailed architecture of a subnetwork (top) with the structure of a convolutional block (bottom left) and fully connected block (bottom right). Filter size ($m$) and number ($n$) for the convolutional layer and output size ($k$) for the fully connected layer are specified.

The detailed architecture of VTPN in this study is displayed in Fig. 1(b)-(c). We chose the DVH of a plan as the input to the VTPN, as it is usually the starting point that a human planner uses to evaluate the plan quality. The input had three columns corresponding to the DVHs of PTV and two organs at risk (OARs). Considering that there were five TPPs to adjust, VTPN was constructed to contain five subnetworks, each designed for one TPP. For each one, we considered three possible adjustment actions: changing the TPP by a factor of $\alpha$ ($\alpha > 1$), $1/\alpha$, or keeping it unchanged. We chose $\alpha = e^{0.5}$ in this study empirically, as we expect the choice of this value would not critically affect the parameter-adjustment performance, but only the speed to reach convergence.

### 2.3 Plan quality evaluation metric

A plan quality evaluation metric $\Phi(\cdot)$ was needed, so that the reward function in DRL can be defined to quantitatively assess the plan quality change caused by TPP adjustments. The VTPN can then learn a policy to maximize $\Phi(\cdot)$. Similar to (Shen *et al.*, 2020a), we employed the PlanIQ score (ProKnow Systems, Sanford, FL, USA) for prostate IMRT as the evaluation metric. The scoring system consisted of a set of criteria to evaluate plan quality based on target coverage and dose conformity, as well as sparing of OARs. For each criterion, a score was defined as a piecewise linear function ranging between 0 and 1. The final plan score was computed as the summation of the scores for all the criteria. A higher score indicated a better plan quality. In this study, we removed the score evaluating PTV underdosage, since such a requirement was enforced by the optimization model (1) as a hard constraint, and therefore held for all the plans. The rest of the criteria we considered included one for PTV overdosage, four criteria for bladder, and another four criteria for rectum, see Table 1. As a consequence, the score for PTV was within the range of [0, 1] ($\text{PlanIQ}_{\text{PTV}}(s) \in [0,1]$), while the scores for bladder and rectum were each within [0, 4] ($\text{PlanIQ}_{\text{BLA}}(s) \in [0,4]$, and $\text{PlanIQ}_{\text{REC}}(s) \in [0,4]$), where $s$ denotes the DVH of a plan. The highest achievable score for a plan was 9.

**Table 1.** Criteria employed in the PlanIQ scoring system for plan quality evaluation.

| Quantity of interest | Scoring Criterion |
|---|---|
| PTV D[0.03cc] (Gy) | $\text{Score} = \begin{cases} 1, & \text{if PTV D[0.03cc]} < 84.4\text{Gy} \\ \frac{\text{PTV D[0.03cc]} - 87.12\text{ Gy}}{84.4\text{ Gy} - 87.12\text{ Gy}}, & \text{if } 84.4\text{Gy} \leq \text{PTV D[0.03cc]} \leq 87.12\text{Gy} \\ 0, & \text{if PTV D[0.03cc]} > 87.12\text{Gy} \end{cases}$ |
| Bladder V[80Gy] (%) | $\text{Score} = \begin{cases} 1, & \text{if Bladder V[80Gy]} < 15\% \\ \frac{\text{Bladder V[80Gy]} - 20\%}{15\% - 20\%}, & \text{if } 15\% \leq \text{Bladder V[80Gy]} \leq 20\% \\ 0, & \text{if Bladder V[80Gy]} > 20\% \end{cases}$ |
| Bladder V[75Gy] (%) | $\text{Score} = \begin{cases} 1, & \text{if Bladder V[75Gy]} < 25\% \\ \frac{\text{Bladder V[75Gy]} - 30\%}{25\% - 30\%}, & \text{if } 25\% \leq \text{Bladder V[75Gy]} \leq 30\% \\ 0, & \text{if Bladder V[75Gy]} > 30\% \end{cases}$ |
| Bladder V[70Gy] (%) | $\text{Score} = \begin{cases} 1, & \text{if Bladder V[70Gy]} < 35\% \\ \frac{\text{Bladder V[70Gy]} - 40\%}{35\% - 40\%}, & \text{if } 35\% \leq \text{Bladder V[70Gy]} \leq 40\% \\ 0, & \text{if Bladder V[70Gy]} > 40\% \end{cases}$ |
| Bladder V[65Gy] (%) | $\text{Score} = \begin{cases} 1, & \text{if Bladder V[65Gy]} < 50\% \\ \frac{\text{Bladder V[65Gy]} - 55\%}{50\% - 55\%}, & \text{if } 50\% \leq \text{Bladder V[65Gy]} \leq 55\% \\ 0, & \text{if Bladder V[65Gy]} > 55\% \end{cases}$ |
| Rectum V[75Gy] (%) | $\text{Score} = \begin{cases} 1, & \text{if Rectum V[75Gy]} < 15\% \\ \frac{\text{Rectum V[75Gy]} - 20\%}{15\% - 20\%}, & \text{if } 15\% \leq \text{Rectum V[75Gy]} \leq 20\% \\ 0, & \text{if Rectum V[75Gy]} > 20\% \end{cases}$ |
| Rectum V[70Gy] (%) | $\text{Score} = \begin{cases} 1, & \text{if Rectum V[70Gy]} < 25\% \\ \frac{\text{Rectum V[70Gy]} - 30\%}{25\% - 30\%}, & \text{if } 25\% \leq \text{Rectum V[70Gy]} \leq 30\% \\ 0, & \text{if Rectum V[70Gy]} > 30\% \end{cases}$ |
| Rectum V[65Gy] (%) | $\text{Score} = \begin{cases} 1, & \text{if Rectum V[65Gy]} < 35\% \\ \frac{\text{Rectum V[65Gy]} - 40\%}{35\% - 40\%}, & \text{if } 35\% \leq \text{Rectum V[65Gy]} \leq 40\% \\ 0, & \text{if Rectum V[65Gy]} > 40\% \end{cases}$ |
| Rectum V[60Gy] (%) | $\text{Score} = \begin{cases} 1, & \text{if Rectum V[60Gy]} < 50\% \\ \frac{\text{Rectum V[60Gy]} - 55\%}{50\% - 55\%}, & \text{if } 50\% \leq \text{Rectum V[60Gy]} \leq 55\% \\ 0, & \text{if Rectum V[60Gy]} > 55\% \end{cases}$ |

*2.4 Training the virtual treatment planner network*

*2.4.1 Standard deep reinforcement learning process*

Before introducing the proposed KgDRL, we will first briefly review the standard DRL framework. The end-to-end DRL training process is derived based on Bellman equation (Bellman and Karush, 1964), a general property of the optimal action-value function. Let $Q(s, a; \theta)$ denotes the VTPN. $\theta$ indicates the network parameters to be determined via the training process. $s$ and $a$ are the DVH of an optimized plan under a given set of TPPs and a TPP adjustment action, respectively. The Bellman equation is:

$$Q(s, a; \theta) = r + \gamma \max_{a'} Q(s', a'; \theta). \tag{2}$$

$s'$ indicates the DVH of the plan obtained after solving the optimization problem with the TPPs updated by applying the action $a$. $r$ is the reward function for the action $a$ acting on the state $s$ to generate the state $s'$. It was computed as the change in plan quality score comparing $s$ and $s'$. Under such a formulation, $Q(s, a; \theta)$ predicts the gain in plan quality associated with the action $a$ for the input DVH of the plan. The ultimate goal of DRL is to build a VTPN satisfying the Bellman equation. Hence, the training process can be simply formulated as minimizing the deviation of $Q(s, a; \theta)$ from the Bellman equation with respect to $\theta$:

$$\min_{\theta} \left[ r + \gamma \max_{a'} Q(s', a'; \theta) - Q(s, a; \theta) \right]^2. \tag{3}$$

Specifically for the standard DRL process using the $\epsilon$-greedy algorithm, starting with a state $s$ obtained using an initial set of TPPs, with probability of $\epsilon$, the DRL process took a random action $a$ among all the possible actions, while chose $a = \mathrm{argmax}_a Q(s, a; \theta)$ with a probability of $(1 - \epsilon)$. After that, $s'$ was obtained by solving the optimization problem using the updated TPPs, while $r$ was computed by comparing the quality of $s'$ with $s$ using the plan scoring system, i.e. $r = \Phi(s') - \Phi(s)$. Repeating such an $\epsilon$-greedy search process and recording all the state-action pairs generated a pool of training samples $\{s, a, s', r\}$. During this process, a strategy called experience replay was performed, which solved the optimization problem in Eq. (3) and updated $\theta$ using samples randomly picked from the pool of training data. Randomly selecting samples prevented training from being affected by the correlation among sequentially generated actions and plans. In this process, VTPN learnt the consequences of applying different TPP adjustment actions to a large number of plans, and the optimal TPP adjustment policy can be gradually identified. This standard DRL algorithm is summarized in Algorithm 1, which was successfully applied to realize intelligent automatic treatment planning in our previous studies (Shen *et al.*, 2019a; Shen *et al.*, 2020a).

---

**Algorithm 1.** Standard DRL algorithm to train VTPN.

    1.   Initialize network coefficients $\theta$;

   **for** episode $= 1, 2, \dots, N_{episode}$

      **for** $k = 1, 2, \dots, N_{patient}$ **do**

        2. Initialize $\lambda, \lambda_{bla}, \lambda_{rec}, \tau_{bla}, \tau_{rec}$

        Solve optimization problem (1) with $\{\lambda, \lambda_{bla}, \lambda_{rec}, \tau_{bla}, \tau_{rec}\}$ for $s^1$;

---

**for** $l = 1, 2, \ldots, N_{train}$ **do**

3. Select an action $a^i$ with $\epsilon$-greedy:

    **Case 1:** with probability $\epsilon$, select $a^i$ randomly;

    **Case 2:** otherwise $a^i = \arg\max_a Q(s^l, a; \theta)$;

4. Update TPPs using $a^i$;

5. Solve optimization problem (1) with updated TPPs for $s^{l+1}$;

6. Compute reward $r^l = \Phi(s^{l+1}) - \Phi(s^l)$;

7. Store state-action pair $\{s^l, a^l, r^l, s^{l+1}\}$ in training data pool;

8. Train $\theta$ with experience replay:

    Randomly select $N_{batch}$ training data from training data pool;

    Update $\theta$ using gradient descent algorithm to solve (3);

    **end for**

    **end for**

**end for**

**Output** $\theta$

---

### 2.4.2 Human knowledge in treatment planning parameter adjustment

The key to improve the training efficiency was to effectively guide the navigation in the state-action space, so the training process can observe more actions that can improve plan quality than the standard DRL training process using the $\epsilon$-greedy algorithm. As such, we proposed to integrate human knowledge in TPP adjustment with the $\epsilon$ -greedy algorithm.

It is an important question to what extent the human knowledge on TPP adjustment would cover. On one hand, it is necessary to define comprehensive rules to effectively guide the DRL training process. On the other hand, rules may not be perfect, as they are concluded based on human experience. The rules should cover only a limited number of scenarios, so that the DRL training can still freely explore the state-action space to discover TPP adjustment policy that is beyond the defined rules. In this proof-of-principle study, we considered three scenarios in the human knowledge set, one for each of PTV, bladder, and rectum. More specifically, let

$$C_H = \{s | \text{PlanIQ}_{\text{PTV}}(s) \leq 0.5, \text{ or PlanIQ}_{\text{BLA}}(s) \leq 2, \text{ or PlanIQ}_{\text{REC}}(s) \leq 2\}, \quad (4)$$

represent the set containing three scenarios in which the PlanIQ score of $\text{PlanIQ}_{\text{PTV}}$, $\text{PlanIQ}_{\text{BLA}}$, and $\text{PlanIQ}_{\text{REC}}$ for PTV, bladder, and rectum are lower than threshold values, i.e. 50% of the corresponding maximal achievable scores (maximal score of 1 for PTV, and 4 for bladder and rectum). $s \in C_H$ indicates that the plan falls into one of the scenarios. $P_H(s)$ is an action defined based on human planner's experience to adjust the TPPs. Based on the clinical importance, as well as our experience with the in-house developed TPS, we considered the following rules $P_H(\cdot)$ with the highest to the lowest priorities:

**Rule 1:** If $\text{PlanIQ}_{\text{PTV}}(s) \leq 0.5$, increase the value of $\lambda$ by $\alpha$;

**Rule 2:** Else if $\text{PlanIQ}_{\text{BLA}}(s) \leq 2$ and $\text{PlanIQ}_{\text{BLA}}(s) \leq \text{PlanIQ}_{\text{REC}}(s)$,

**Case 1:** with probability $\epsilon_{BLA}$, increase $\lambda_{BLA}$ by $\alpha$;

**Case 2:** otherwise, decrease $\tau_{BLA}$ by $\alpha$;

**Rule 3:** Else if $\text{PlanIQ}_{REC}(s) \leq 2$ and $\text{PlanIQ}_{REC}(s) \leq \text{PlanIQ}_{BLA}(s)$,

**Case 1:** with probability $\epsilon_{REC}$, increase $\lambda_{REC}$ by $\alpha$;

**Case 2:** otherwise, decrease $\tau_{REC}$ by $\alpha$;

Note that PTV had the highest priority among all the three rules. Once the score of PTV was lower than the threshold, the rule would adjust the parameter $\lambda$ to reduce PTV overdose and improve the score. The priority between bladder and rectum depends on the specific situation. The rules were set to always adjust TPPs for the organ with a lower score among the two organs.

### 2.4.3 Knowledge-guided deep reinforcement learning

The proposed KgDRL employed the aforementioned human rule as an additional brunch to the $\epsilon$-greedy algorithm (Fig. 2), letting the rule to guide the training process of the VTPN. More specifically, given a state $s$, if $s \in C_H$, the TPP adjustment action $a$ was determined by the rules with a large probability $\epsilon_H$, i.e. $a = P_H(s)$, while the action was determined via the standard $\epsilon$-greedy algorithm otherwise. Note that this additional choice did not largely increase the computational complexities from DRL in each episode, as majority of the computation costs were spent on solving the plan optimization problem based on determined TPPs, not on how to determine the actions. By involving $P_H(\cdot)$ as the guidance to navigate the state-action space, we hoped to identify the proper TPP adjustment policy to improve the plan quality with much less training steps than the standard DRL training process. The workflow of the KgDRL algorithm is shown in Fig. 2 and outlined in Algorithm 2 in detail.

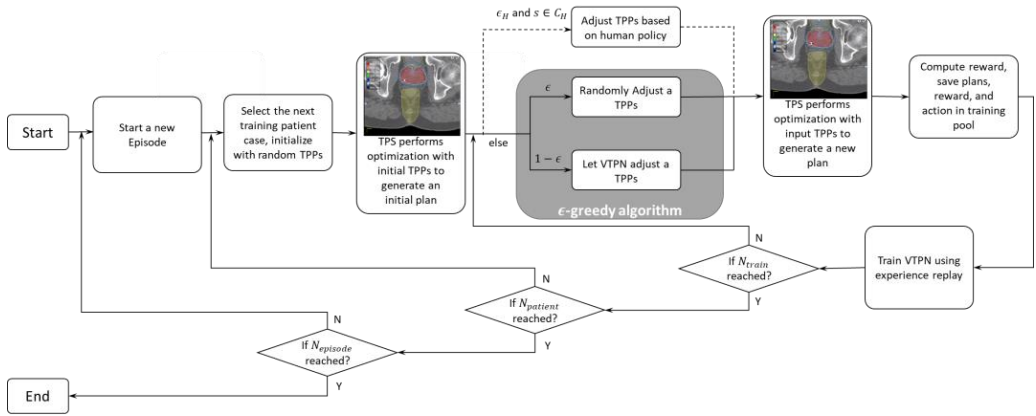

**Figure 2.** Integrating human TPP adjustment rules with the $\epsilon$-greedy algorithm to generate training samples in the DRL process. Such a process reduces to the standard DRL, if removing the dashed brunch corresponding to the incorporation of rules.

---

**Algorithm 2.** KgDRL algorithm to train VTPN.

---

1. Initialize network coefficients $\theta$;

**for** episode $= 1, 2, \ldots, N_{episode}$

    **for** $k = 1, 2, \ldots, N_{patient}$ **do**

        2. Initialize $\lambda, \lambda_{bla}, \lambda_{rec}, \tau_{bla}, \tau_{rec}$

          Solve optimization problem (1) with $\{\lambda, \lambda_{bla}, \lambda_{rec}, \tau_{bla}, \tau_{rec}\}$ for $s^1$;

        **for** $l = 1, 2, \ldots, N_{train}$ **do**

            3. Select an action $a^l$ based on human rule or $\epsilon$-greedy:

                **Case 1:** with probability $\epsilon_H$, if $s \in C_H$, $a^l = P_H(s^l)$;

                **Case 2:** otherwise apply $\epsilon$-greedy algorithm:

                    **Case 2.1:** with probability $\epsilon$, select $a^l$ randomly;

                    **Case 2.2:** otherwise $a^l = \arg\max_a Q(s^l, a; \theta)$;

            4. Update TPPs using $a^l$;

            5. Solve optimization problem (1) with updated TPPs for $s^{l+1}$;

            6. Compute reward $r^l = \Phi(s^{l+1}) - \Phi(s^l)$;

            7. Store state-action pair $\{s^l, a^l, r^l, s^{l+1}\}$ in training data pool;

            8. Train $\theta$ with experience replay:

              Randomly select $N_{batch}$ training data from training data pool;

              Update $\theta$ using gradient descent algorithm to solve (3);

        **end for**

        **end for**

    **end for**

    **Output** $\theta$

---

### 2.5 Implementation details and evaluations

We collected 74 patient cases with prostate cancer treated with IMRT. Aside from the 10 patient cases randomly picked to train the VTPNs, we also randomly selected 5 patients for validation purpose. The remaining 59 patient cases were saved for testing. We trained the VTPN using the proposed KgDRL approach, as well as another VTPN with the standard DRL algorithm (Shen *et al.*, 2020a), i.e. Algorithm 1 for the comparison purpose. The network architectures and experimental setups for both VTPNs were identical for a fair comparison. The training step $N_{train}$ was set to 30. For each patient case, we started with all TPPs that were set to be unity. The initial probability $\epsilon_H$ was 0.7 in KgDRL and $\epsilon$ for $\epsilon$-greedy algorithm for both cases was set to be 0.99. $\epsilon_H$ and $\epsilon$ decreased with the same rate of 0.99 per episode over the training process. In addition, $\epsilon_{BLA}$ and $\epsilon_{REC}$, i.e. the probabilities defined for human rules were set to 0.2 based on experience from human planners.

All the computations was performed using Python with TensorFlow (Abadi *et al.*, 2016) on a desktop workstation with eight Intel Xeon 3.5 GHz CPU processors, 32 GB memory and two Nvidia Quadro M4000 GPU cards.

After successfully training the VTPN using the standard DRL and the KgDRL, we evaluated their performances in 59 patient cases that were not seen in the training step in three setups. 1) Rule-based planning. For each case, we first set all TPPs to unity to generate an initial plan. After that, human rules summarized in section 2.3.2 were utilized to repeatedly adjust TPPs until $s \notin C_H$. The purpose of this experiment was to investigate the effectiveness of the summarized rules in treatment planning. 2) Planning using VTPN trained with KgDRL. We first initialized all TPPs to unity and then employed the VTPN trained with KgDRL to adjust TPPs and generate a plan. The iteration of TPP adjustment was continued, until one of the following three criteria was met: the plan reached the maximal score of 9, VTPN decided to keep all TPPs unchanged, or a maximal number of adjustment steps (50) was reached. 3) Planning using VTPN trained with the standard DRL. This was the same as in 2) except that VTPN trained with the standard DRL was employed.

## 3. Results

### 3.1.1 Training efficiency and effectiveness of DRL and KgDRL

The rewards and Q-values along the training episodes for VTPNs trained via KgDRL and DRL are shown in Fig. 3(a) and (b). Fig. 3(c) are PlanIQ score as a function of the training episode, which was computed using the VTPN trained to the episode number to plan the cases in the validation patient dataset. Rewards reflect the improvement in plan quality obtained via automatic TPP adjustment using the VTPN, while the Q-values indicate the output of VTPNs. The higher these values are, it is expected that the VTPN's performance is better.

In general, an increasing trend in rewards and Q-values along the training process was observed for both DRL and KgDRL, illustrating the effectiveness of both training schemes. However, the reward and Q-value of KgDRL increased much faster compared to the standard DRL. For instance, we observed that the reward and Q-value of KgDRL at the 8-th episodes were approximately the same to those of DRL at the 100-th episode. Note that we stopped the KgDRL at the 10-th episode, as we observed satisfactory performance on the validation data at the 8-th episode, and the average plan score was not improved afterwards, see Fig. 3(c). On the other hand, DRL improved the average score of VTPN generated plans gradually, until it achieved satisfactory performance after 100 episodes. This comparison indicated that it took KgDRL only 8% of episodes required by DRL to reach the same level of intelligence in VTPN. Since the computational complexities for DRL and KgDRL in each training episode were similar, the reduction in the number of episodes translated immediately to the reduction in computation time. It took ~13 hours for KgDRL to complete the training process, as compared to approximately a week time for DRL.

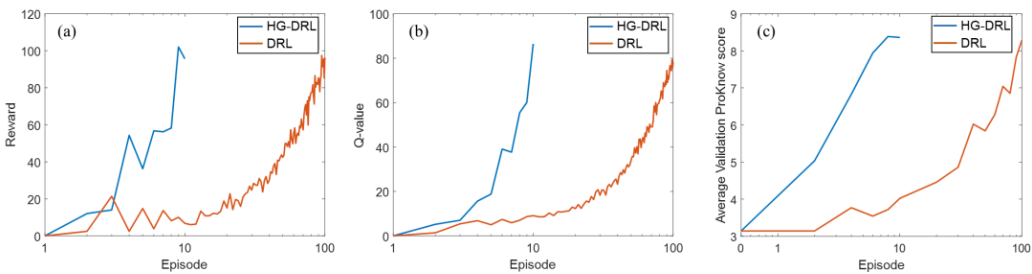

**Figure 3.** Comparison of rewards (a) and Q-values (b), and validation performance (c) along training episodes between KgDRL and DRL.

### 3.1.2  Testing performance of VTPNs trained via KgDRL

In this section, we studied the treatment planning performance using purely the human rules, and the two VTPNs trained via KgDRL with 8 episodes and DRL with 100 episodes, respectively. In addition, we also studied the performance of the VTPN trained via 8 episodes of the standard DRL to further highlight the training efficiency gain of KgDRL.

In Fig. 4, we first show the treatment planning process for one representative patient case performed by the VTPN trained with 8 episodes of KgDRL. It was observed that the VTPN successfully improved the plan quality, as evidenced by the generally increasing trend of the plan score along the planning process, and finally reaching the maximal score of 9. Specifically, the VTPN firstly determined to decrease the value of $\tau_{rec}$ to improve rectum sparing, and then focused on eliminating hot spots and enhancing the PTV dose homogeneity by raising $\lambda$, the weighting factor of PTV overdose in the objective function. It then changes the weighting factors of rectum and bladder, respectively, to adjust their importance. Later on, $\tau_{bla}$ was adjusted to reduce the dose delivered to bladder. In 18 steps of TPP adjustment, the plan score reached 9, i.e. the highest score in our scoring system, which concluded the treatment planning process. The planning process using VTPN trained with 100 episodes of the standard DRL was similar, and hence is not presented.

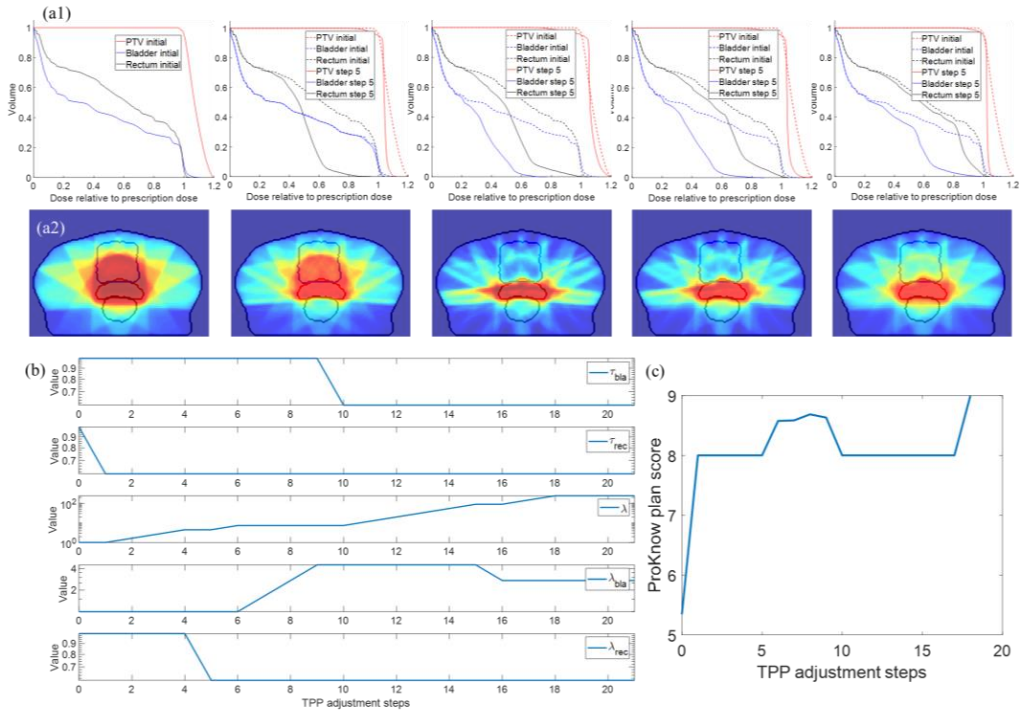

**Figure 4.** Evolution of DVH, dose distribution, TPPs, and PlanIQ scores for a representative testing patient case in the planning process performed by the VTPN trained with 8 episodes of KgDRL. (a1) From left to right: DVHs at TPP adjustment steps 0 (initial plan), 5, 10, 15, and 18 (final step) compared with that of the initial plan. (a2) Corresponding dose distributions. (b) and (c) TPP values and PlanIQ plan scores along the planning process.

We evaluated the effectiveness of using human rules for TPP adjustment on all testing patients, see Fig. 5 and Table 2. Compared to the initial plans that were generated with all TPPs set to unity, the rules were able to improve the plan score from 4.97 ($\pm$2.02) to 7.81 ($\pm$1.59) (average plan score ($\pm$ standard deviation)). Yet, there was still rooms to further improve the resulting plans, especially for those receiving relatively low scores. The main reason for the relatively low performance of the rule-based planning process was that we only selected three general rules being valid for most of the patient cases. These rules were not complete and by no means optimal for each specific patient. In particular, the rules failed to improve plans for two of the testing patient cases, which made the resulting plan scores less than one. This in fact highlighted the need of developing the VTPN to learn how to intelligently adjust TPPs for specific patient cases.

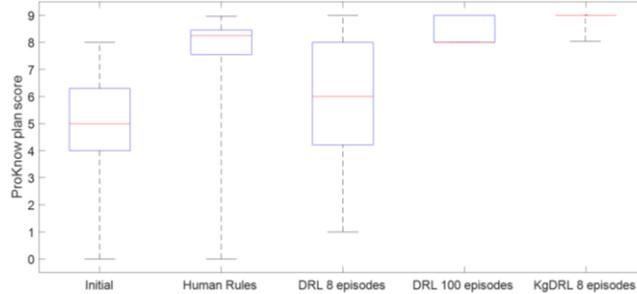

**Figure 5.** A box plot comparing PlanIQ plan scores. Each blue box covers the 25th and 75th percentiles of the plan scores on 59 testing patient cases, while the red line in the middle gives the median value. Top and bottom lines indicate maximal and minimal scores. From left to right: initial plans, plans generated using only rules, plans generated by VTPN trained with 8 episodes of DRL, plans generated by VTPN trained with 100 episodes of DRL, and plans generated by VTPN trained with 8 episodes of KgDRL.

**Table 2.** Comparison of performance using rules only, DRL and KgDRL on testing dataset.

|  | Initial | Rules | DRL | | KgDRL |
|---|---|---|---|---|---|
| Number of training episodes | -- | -- | 8 | 100 | 8 |
| Training time (hours) | -- | **--** | 13 | 172 | 13 |
| Average PlanIQ score ($\pm$ standard deviation) | 4.97 ($\pm$2.02) | 7.81 ($\pm$1.59) | 5.87 ($\pm$2.37) | 8.43 ($\pm$0.48) | 8.82 ($\pm$0.29) |

Compared to purely relying on rules, both VTPNs trained via KgDRL and DRL were able to achieve better performance. More specifically VTPN trained with only 8 episodes of KgDRL achieved an average plan score of 8.82 on all the testing patient cases, with most of the testing cases (48 out of 59) reached the maximal plan score of 9. With a similar performance level, the VTPN trained with 100 episodes of DRL was capable of reaching an average score of 8.43. Plan scores of all the testing cases were at least 8, higher than the average plan score achieved by using the rules only. Based on these results, VTPNs were successfully trained with DRL using 100 episodes and KgDRL using 8 episodes, demonstrating the effectiveness of the DRL framework in learning TPP-adjustment policy for high-quality treatment plans.

The advantage of KgDRL can also be observed from the angle of comparing the performance of VTPNs trained with KgDRL and DRL, but both with 8 episodes. In this case, the standard DRL was not able to fully train the VTPN to a proficient level in treatment planning. The resulting VTPN was only able to improve the average PlanIQ plan score from 4.97 to 5.87, much lower than the VPTN trained with KgDRL using the same number of episodes.

## 4.  Discussions

The current study introduced an effective approach to incorporate human knowledge in the process of training a VTPN. Tests demonstrated that this approach was able to substantially improve the training efficiency. Although the current study focused on an exemplary problem of treatment planning for prostate cancer IMRT using an in-house TPS, the achieved success was expected to be of vital importance for the development of the DRL-based intelligent automatic treatment planning framework towards more complicated tumor sites, e.g. head and neck (H&N) cancer, more complex treatment planning problems, e.g. VMAT, and the incorporation of clinically realistic TPSs with more options of TPP adjustments. In these cases, the much larger state-action space and longer time to solve a plan optimization problem would significantly increase the computational challenge to train a VTPN, which could make the standard DRL training approach impractical. The proposed KgDRL approach with significantly improved efficiency offers a potential solution to this problem.

It is worthwhile to discuss to what extent we should define rules and incorporate them in the training process.  Generally speaking, as illustrated in Fig. 6, in the large space spanned by states and actions, the optimal policy to operate the TPS for treatment planning is represented as an action function of the state, i.e. deciding an action for a state. The process of training a VTPN is essentially to find an approximation using the deep neural network to represent this function. This function apparently occupies only a small portion of the entire state-action space. The ineffectiveness of the conventional DRL using the $\epsilon$-greedy approach comes from the random exploration in the large state-action space, and hopefully being able to observe enough training data falling in the space of the optimal policy to allow training the VTPN. The proposed KgDRL approach, instead, defines a small space of rules (Fig. 6) to help generating training data within the desired region, hence accelerating the training process.

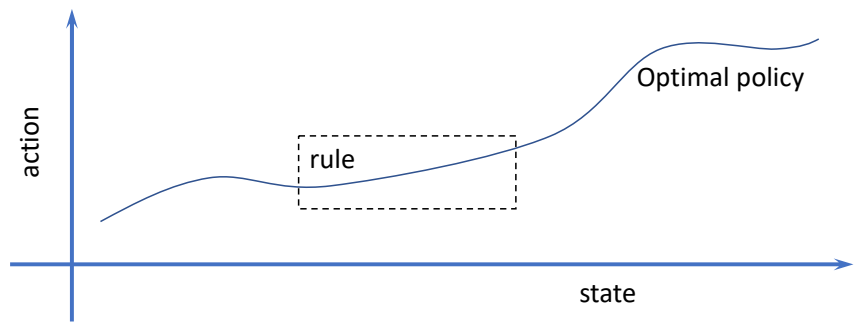

**Figure 6.** Illustration of the relationship between the entire state-action space, optimal policy space, and rule space.

The key to the success of KgDRL is effectiveness and generality of rules. Effectiveness means that the rules can improve plan quality, when using them to generate actions in treatment planning. Generality means that the effectiveness of rules can be realized on a large population of patients. Those rules satisfying both conditions can significantly increase the chance for a VTPN to observe proper ways of TPP adjustment, and learn from them, such that the training of the VTPN can be made more efficiently than training without using any guidance. In practice, it may be straightforward to define effective rules for a certain number of simple scenarios, such as the three ones considered in this study. However, it would be challenging to define rules that are both effective and general, as patient cases are different, and patient-specific rules may be required. One may introduce a set of very complex rules that can produce high-quality plans for each specific scenario one could think of. Yet doing so is not only tedious, but it is hard to ensure the validity of rules in unseen scenarios and hence likely to sacrifice the generality of the rules. Using these rules to guide the development of VTPN would risk at problems of misleading the training process.

One excitement observed in this study was that it seemed the training of VTPN did not require very effective rules. The three very general rules in this study (Sec. 2.4.2) were relatively simple, but not very effective. As demonstrated by the study in Sec. 3.1.2, the performance of treatment planning purely based on rules was not satisfactory, resulting in the final PlanIQ score of 7.81 ($\pm$1.59). This indicated that the rules were only a subset of the optimal policy space. Yet, these rules served as seeds for the VTPN to grow and eventually the training process enabled the VTPN to discover the rest of the optimal policy space. The discovered policy was found to be effective and general for treatment planning, as indicated by the average score of 8.82 ($\pm$0.29) achieved by the VTPN on a number of patients that were not seen in the training process.

Integrating human knowledge with state-of-the-art deep learning techniques is actually a topic of great importance (Choo and Liu, 2018; Zhang *et al.*, 2018; Ning *et al.*, 2017; Bian *et al.*, 2014), as doing so not only helps improving effectiveness of building a deep learning model, it also often enhances other aspects of the model, such as interpretability. The proposed study provided a potential approach, but there are other possible solutions to serve the same purpose. For instance, the proposed KgDRL approach used rules to guide the training of a deep learning model with human experience. The resulting VTPN may or may not agree with the rules in those scenarios that rules were defined. To a certain extent, this approach is similar to having a "soft" constraint in an optimization problem: constraint on the agreement between the rules and the resulting policy discovered by the VTPN. On the other hand, we may also treat rules as "hard" constraints when building the VTPN, if there exist rules that are known for certain effective. In this strategy, we may set up the VTPN to contain two parts, one representing the rules and it does not require training, and one representing the remaining optimal policy space complement to the defined rules. Furthermore, it may be even possible that building a clinically applicable VTPN may rely on both the soft and the hard constraint approaches. It will be our future work to explore further along this direction.

The current study has several limitations. First, similar to our previous study (Shen *et al.*, 2019a; Shen *et al.*, 2020a), the reward function derived from the PlanIQ score may not fully represent the clinical objectives in treatment planning. The current study focused on improving the efficiency of training VTPN by incorporating human knowledge, and the validity of PlanIQ score is beyond this scope. However, to ensure future clinical applicability of VTPN, it is of central importance to model the criteria of more clinical relevance, such as physician's judgement, as reward function to build a VTPN with a practical value for clinical practice. Motivated by the recent advancements in inverse deep reinforcement learning (Wulfmeier *et al.*, 2015) that allowed learning the reward function based on human behaviors, we plan to learn and incorporate the physician's preference into the training of the VTPN. Second, the current study can only serve as a proof-of-principle one to demonstrate the effectiveness of VTPN in a simplified treatment planning problem using an in-house TPS. Extending the VTPN with a clinically realistic TPS is currently an on-going work in our group, and we will report our process in future publications. Last, but not the least, another limitation of our approach was the simple network structure. Under the current formulation (Fig. 1), the size of the network would increase linearly with respect to the number of TPPs involved in the plan optimization problem. As we can imagine, the network size of a VTPN to automatically operate a real TPS for the treatment planning of more complicate cancer sites would be huge. This would pose substantial challenge in computations. Hence, improving the scalability of the VTPN is needed and further study will be down the road.

## 5. Conclusions

In this paper, we have proposed an KgDRL framework to integrate human experience with DRL for intelligent automatic treatment planning. Using prostate cancer IMRT treatment planning as a testbed, we showed that rules defined based on human experience was able to effectively guide the navigation process in the large state-action space. which substantially improved the training efficiency of VTPN. Compared to the standard DRL training approach using the $\epsilon$-greedy algorithm, KgDRL reduced the training time by over 90%. The efficiency gain of KgDRL would potentially enable the applications of DRL to complicated but more clinically relevant treatment planning problems. This study also showed a successful example of employing human knowledge to enhance the state-of-the-art deep learning techniques.

**Acknowledgement**

This work was supported by the National Institutes of Health grant number R01CA237269.

## References

Abadi M, Barham P, Chen J, Chen Z, Davis A, Dean J, Devin M, Ghemawat S, Irving G and Isard M 2016 TensorFlow: A System for Large-Scale Machine Learning. In: *OSDI*, pp 265-83

Bellman R and Karush R 1964 Dynamic programming: a bibliography of theory and application. RAND CORP SANTA MONICA CA)

Bian J, Gao B and Liu T-Y *Joint European conference on machine learning and knowledge discovery in databases,2014),* vol. Series): Springer) pp 132-48

Boutilier J J, Lee T, Craig T, Sharpe M B and Chan T C Y 2015 Models for predicting objective function weights in prostate cancer IMRT *Medical Physics* **42** 1586-95

Boyd S, Parikh N, Chu E, Peleato B and Eckstein J 2011 Distributed optimization and statistical learning via the alternating direction method of multipliers *Foundations and Trends® in Machine Learning* **3** 1-122

Chan T C Y, Craig T, Lee T and Sharpe M B 2014 Generalized Inverse Multiobjective Optimization with Application to Cancer Therapy *Operations Research* **62** 680-95

Choo J and Liu S 2018 Visual Analytics for Explainable Deep Learning *IEEE Computer Graphics and Applications* **38** 84-92

Das I J, Cheng C-W, Chopra K L, Mitra R K, Srivastava S P and Glatstein E 2008 Intensity-Modulated Radiation Therapy Dose Prescription, Recording, and Delivery: Patterns of Variability Among Institutions and Treatment Planning Systems *JNCI: Journal of the National Cancer Institute* **100** 300-7

Fan J, Wang J, Chen Z, Hu C, Zhang Z and Hu W 2019 Automatic treatment planning based on three‐dimensional dose distribution predicted from deep learning technique *Med Phys* **46** 370-81

Glowinski R and Le Tallec P 1989 *Augmented Lagrangian and operator-splitting methods in nonlinear mechanics* vol 9: SIAM)

Holdsworth C, Kim M, Liao J and Phillips M 2012 The use of a multiobjective evolutionary algorithm to increase flexibility in the search for better IMRT plans *Med Phys* **39** 2261-74

Holdsworth C, Kim M, Liao J and Phillips M H 2010 A hierarchical evolutionary algorithm for multiobjective optimization in IMRT *Med Phys* **37** 4986-97

Lee T, Hammad M, Chan T C Y, Craig T and Sharpe M B 2013 Predicting objective function weights from patient anatomy in prostate IMRT treatment planning *Medical Physics* **40** 121706-n/a

Lu R, Radke R J, Happersett L, Yang J, Chui C-S, Yorke E and Jackson A 2007 Reduced-order parameter optimization for simplifying prostate IMRT planning *Physics in Medicine & Biology* **52** 849

Mahmood R, Babier A, McNiven A, Diamant A and Chan T C 2018 Automated treatment planning in radiation therapy using generative adversarial networks *arXiv preprint arXiv:1807.06489*

Mnih V, Kavukcuoglu K, Silver D, Graves A, Antonoglou I, Wierstra D and Riedmiller M 2013 Playing atari with deep reinforcement learning *arXiv preprint arXiv:1312.5602*

Mnih V, Kavukcuoglu K, Silver D, Rusu A A, Veness J, Bellemare M G, Graves A, Riedmiller M, Fidjeland A K, Ostrovski G, Petersen S, Beattie C, Sadik A, Antonoglou I, King H, Kumaran D, Wierstra D, Legg S and Hassabis D 2015 Human-level control through deep reinforcement learning *Nature* **518** 529-33

Nelms B E, Robinson G, Markham J, Velasco K, Boyd S, Narayan S, Wheeler J and Sobczak M L 2012 Variation in external beam treatment plan quality: An inter-institutional study of planners and planning systems *Practical Radiation Oncology* **2** 296-305

Nguyen D, Barkousaraie A S, Shen C, Jia X and Jiang S 2019 Generating Pareto optimal dose distributions for radiation therapy treatment planning *arXiv preprint arXiv:1906.04778*

Nguyen D, McBeth R, Sadeghnejad Barkousaraie A, Bohara G, Shen C, Jia X and Jiang S 2020 Incorporating human and learned domain knowledge into training deep neural networks: A differentiable dose-volume histogram and adversarial inspired framework for generating Pareto optimal dose distributions in radiation therapy *Med Phys* **47** 837-49

Ning G, Zhang Z and He Z 2017 Knowledge-guided deep fractal neural networks for human pose estimation *IEEE Transactions on Multimedia* **20** 1246-59

Shen C, Gonzalez Y, Chen L, Jiang S and Jia X 2018 Intelligent Parameter Tuning in Optimization-Based Iterative CT Reconstruction via Deep Reinforcement Learning *IEEE transactions on medical imaging* **37** 1430-9

Shen C, Gonzalez Y, Klages P, Qin N, Jung H, Chen L, Nguyen D, Jiang S B and Jia X 2019a Intelligent inverse treatment planning via deep reinforcement learning, a proof-of-principle study in high dose-rate brachytherapy for cervical cancer *Physics in Medicine & Biology* **64** 115013

Shen C, Nguyen D, Chen L, Gonzalez Y, McBeth R, Qin N, Jiang S B and Jia X 2020a Operating a treatment planning system using a deep-reinforcement learning-based virtual treatment planner for prostate cancer intensity-modulated radiation therapy treatment planning *Med Phys* **n/a**

Shen C, Nguyen D, Zhou Z, Jiang S B, Dong B and Jia X 2020b An introduction to deep learning in medical physics: advantages, potential, and challenges *Physics in Medicine & Biology* **65** 05TR1

Shen C, Tsai M-Y, Gonzalez Y, Chen L, Jiang S B and Jia X *15th International Meeting on Fully Three-Dimensional Image Reconstruction in Radiology and Nuclear Medicine,2019b),* vol. Series 11072): International Society for Optics and Photonics) p 1107203

Silver D, Huang A, Maddison C J, Guez A, Sifre L, Van Den Driessche G, Schrittwieser J, Antonoglou I, Panneershelvam V and Lanctot M 2016 Mastering the game of Go with deep neural networks and tree search *nature* **529** 484-9

Silver D, Schrittwieser J, Simonyan K, Antonoglou I, Huang A, Guez A, Hubert T, Baker L, Lai M and Bolton A 2017 Mastering the game of Go without human knowledge *Nature* **550** 354-9

Wahl N, Bangert M, Kamerling C P, Ziegenhein P, Bol G H, Raaymakers B W and Oelfke U 2016 Physically constrained voxel-based penalty adaptation for ultra-fast IMRT planning *Journal of Applied Clinical Medical Physics* **17** 172-89

Wang H, Dong P, Liu H and Xing L 2017 Development of an autonomous treatment planning strategy for radiation therapy with effective use of population-based prior data *Medical Physics* **44** 389-96

Watkins C J and Dayan P 1992 Q-learning *Machine learning* **8** 279-92

Wu X and Zhu Y 2001 An optimization method for importance factors and beam weights based on genetic algorithms for radiotherapy treatment planning *Physics in Medicine & Biology* **46** 1085

Wulfmeier M, Ondruska P and Posner I 2015 Maximum entropy deep inverse reinforcement learning *arXiv preprint arXiv:1507.04888*

Xing L, Li J G, Donaldson S, Le Q T and Boyer A L 1999 Optimization of importance factors in inverse planning *Physics in Medicine and Biology* **44** 2525

Yan H and Yin F-F 2008 Application of distance transformation on parameter optimization of inverse planning in intensity-modulated radiation therapy *Journal of Applied Clinical Medical Physics* **9** 30-45

Yan H, Yin F-F, Guan H and Kim J H 2003a Fuzzy logic guided inverse treatment planning *Medical Physics* **30** 2675-85

Yan H, Yin F-F, Guan H-q and Kim J H 2003b AI-guided parameter optimization in inverse treatment planning *Physics in Medicine & Biology* **48** 3565

Yang Y and Xing L 2004 Inverse treatment planning with adaptively evolving voxel-dependent penalty scheme *Medical Physics* **31** 2839-44

Zhang X, Wang S, Liu J and Tao C 2018 Towards improving diagnosis of skin diseases by combining deep neural network and human knowledge *BMC medical informatics and decision making* **18** 59